\documentclass[]{spie}  %>>> use for US letter paper
\usepackage{amsmath,amsfonts,amssymb}
\usepackage{graphicx}
\usepackage[colorlinks=true, allcolors=blue]{hyperref}

\usepackage{comment}
\usepackage[]{graphicx}
\usepackage[]{color}
\usepackage{alltt}

\usepackage{siunitx}
\usepackage{subcaption}
\usepackage{cleveref}
\usepackage{algorithm}
\usepackage{algpseudocode}
\usepackage{makecell}

\algnewcommand\algorithmicto{\textbf{to}}
\algnewcommand\RETURN{\State \textbf{return} }
\usepackage{tikz}
\usetikzlibrary{shapes,arrows,calc,fit}
\usepackage{tikz}
\tikzstyle{block} = [draw, rectangle, 
    minimum height=3em, minimum width=6em]
\tikzstyle{sum} = [draw, circle, node distance=1cm]
\tikzstyle{input} = [coordinate]
\tikzstyle{output} = [coordinate]
\tikzstyle{pinstyle} = [pin edge={to-,thin,black}]
\usepackage{capt-of}
\usepackage[super]{nth}
\providecommand{\keywords}[1]{\textbf{\textit{Keywords:}} #1}
\newcommand\norm[1]{\left\lVert#1\right\rVert}
\title{Upgrading SPHERE with the second stage AO system SAXO+: frequency-based data-driven controller for adaptive optics}

\author[a]{Isaac Dinis}
\author[a]{François Wildi}
\author[a]{Damien Ségransan}
\author[b]{Vaibhav Gupta}
\author[b]{Alireza Karimi}
\author[c]{Michel Tallon}
\author[c]{Isabelle Bosc}
\author[c]{Maud Langlois}
\author[c]{Magali Loupias}
\author[c]{Clémentine Bechet}
\author[c]{Eric Thiébaut}
\author[d]{Charles Goulas}
\author[d]{Florian Ferreira}
\author[d]{Anthony Boccaletti}
\author[d]{Fabrice Vidal}
\author[e]{Caroline Kulcsar}
\author[e]{Henri-François Raynaud}
\author[e]{Nicolas Galland}
\author[f]{Markus Kasper}
\author[g]{Julien Milli}
\author[g]{David Mouillet}
\author[h]{Laura Schreiber}
\author[h]{Emiliano Diolaiti}
\author[h]{Raffaele Gratton}
\author[i]{Gael Chauvin}

\affil[a]{Observatoire de l'université de Genève (UNIGE), Geneva, Switzerland}
\affil[b]{École Polytechnique Fédérale de Lausanne (EPFL), Lausanne, Switzerland}
\affil[c]{Université Claude Bernard Lyon 1 (univ lyon1), Lyon, France}
\affil[d]{Observatoire de Paris (OBSPM), Paris, France}
\affil[e]{Institut d'Optique, Paris, France}
\affil[f]{European Southern Observatory (ESO), Garching, Germany}
\affil[g]{Université Grenoble-Alpes (UGA), Grenoble, France}
\affil[h]{Italian National Institute for Astrophysics (INAF), Bologna, Italy}
\affil[i]{Observatoire de la Côte d'Azur (OCA), Nice, France}

\authorinfo{Further author information: Send correspondence to Isaac Dinis : isaac.dinis@unige.ch}

\begin{document} 
\maketitle

\begin{abstract}
This study introduces a novel frequency-based data-driven controller for adaptive optics, using power spectral density for optimization while ensuring stability criteria. It addresses disturbance rejection, command amplitude constraints and system transfer functions through convex optimization to obtain an optimal control in an infinite input response filter form. Evaluated within the SAXO+\cite{SAXO+} project, it demonstrates efficacy under diverse atmospheric conditions and operational scenarios. The proposed controller is tested in both standard and disentangled adaptive optics schemes \cite{dcao}, showcasing its adaptability and performance. Experimental validation is conducted using the COMPASS\cite{COMPASS} simulation tool, affirming the controller's promise for enhancing adaptive optics systems in real-world applications.
\end{abstract}


\section{INTRODUCTION}
\label{sec:intro}  % \label{} allows reference to this section

Adaptive Optics (AO) is a technology aimed at improving direct imaging for ground-based telescopes by correcting the light wavefront distorted due to atmospheric dispersion. Since its inception, Adaptive Optics has continuously improved, with the most advanced iteration being called eXtreme Adaptive Optics (XAO), exemplified by instruments like the SPHERE \cite{SPHERE} instrument at the VLT. As we enter the era of Extremely Large Telescopes (ELTs), SAXO+\cite{SAXO+} is proposed as a pathfinder for XAO instrument for ELTs, such as PCS\cite{PCS}. SAXO+ aims to enhance SPHERE adaptive optics by adding a second stage AO loop with better hardware \cite{saxo+design}: a faster deformable mirror (DM) capable of running at up to 3 kHz and a more sensitive WaveFront Sensor (WFS) - the Pyramid WFS (PWFS); Advanced control laws are proposed to drive this second loop including LQG \cite{LQG}, TAO\cite{TAO}, reinforcement learning \cite{PO4AO}, or the focus of this paper, the data-driven controller \cite{Gupta}. The data-driven controller can be viewed as an enhancement of the widely used integrator controller. As a reminder, in a discrete system, the command computed by the integrator is equal to a gain times the measurement plus the previous command. Intuitively, one could imagine that if the controller examines further past measurements and commands, it could better predict and compute the new command to apply to reject disturbances. This is known as increasing the controller order, but it has a downside: while it is feasible to manually tune an integrator (only one parameter), it would be impractical to manually tune a fifth-order controller by guessing the gains for the four previous commands and five previous measurements. Fortunately, what cannot be computed by a human can often be computed by a machine nowadays. Therefore, the data-driven controller translates into a convex optimization problem for which plenty of solvers already exist. Three objectives are defined for the optimizer: performance, stroke management, and robustness. The controller is frequency-based, i.e., all three objectives are described with their Power Spectral Density (PSD). To compare and evaluate the different controllers, the SAXO+ control and simulation group designed a simulation framework in COMPASS\cite{COMPASS}. Five science cases were defined to evaluate the performance of the controller for the consolidation phase and are presented here. Additionally, a vibration case, which was not implemented at the time of the consolidation review, is also studied. 

\section{Control theory in adaptive optics}

This section describes the functioning of an AO system using simple block schemes. From the control point of view, an AO system can be resumed with three elements in a closed loop feedback scheme: the WaveFront Sensor (WFS), the Real-Time Computer (RTC) and the Deformable Mirror (DM). 

\begin{comment}
\begin{center}
\makebox[0pt]{%
\begin{tikzpicture}[auto, node distance=2cm,>=latex']

    % We start by placing the blocks
    \node [input, name=input] {};
    \node [sum, right of=input] (sum) {};
    \node [block, right of=sum,node distance=2cm,align=center] (delay) {$WFS$};
    \node [block, right of=delay,node distance=3cm,align=center] (controller) {$RTC$};
    \node [block, right of=controller, node distance=3cm] (dm) {$DM$};

    % We draw an edge between the controller and dm block to 
    % calculate the coordinate u. We need it to place the measurement block. 
    \draw [->] (controller) -- node[name=u] {$u$} (dm);
    \node [output, right of=dm] (output) {};
    \coordinate [below of=u,node distance=1.5cm] (tmp);
    \coordinate [right of=dm] (tmp3);

    % Once the nodes are placed, connecting them is easy. 
    \draw [draw,->] (input) -- node {$d$} (sum);
    \draw [->] (sum) -- node [name=e] {$e$} (delay);
    \draw [->] (delay) -- (controller);
    %\draw [->] (dm) -- node [name=y] {} node [below right] {$-$} (disturbance);
    \draw [->] (dm) -- (tmp3) |- (tmp) -| node[pos=0.99] {} 
        node [pos=0.95] {$-$} (sum);
\end{tikzpicture} 
}
\captionof{figure}{Reference case block diagram, with $d$ the distustbance input, $e$ the residual error and $u$ the command computed by the RTC and applied by the DM.} 
\end{center}
\end{comment}
\begin{center}
\makebox[0pt]{%
\tikzstyle{block} = [draw, rectangle, 
    minimum height=3em, minimum width=6em]
\tikzstyle{sum} = [draw, circle, node distance=1cm]
\tikzstyle{transform_mat} = [draw, isosceles triangle, node distance=0.5cm]
\tikzstyle{input} = [coordinate]
\tikzstyle{output} = [coordinate]
\tikzstyle{mytext} = [coordinate]
\tikzstyle{pinstyle} = [pin edge={to-,thin,black}]
% The block diagram code is probably more verbose than necessary
\begin{tikzpicture}[auto, node distance=2cm,>=latex']
    % We start by placing the blocks
    \node [input, name=input] {};
    \node [sum, right of=input,node distance=1.2cm] (sum) {};
    \node [block, right of=sum,node distance=2.2cm] (wfs) {WFS};
    \node [sum, right of=wfs,node distance=1.6cm,pin={[pinstyle]above:$\vec{n}$}] (sum_noise) {};
    \node [transform_mat, right of=sum_noise,node distance=1.5cm] (s2m) {S2M};
    \node [block, right of=s2m,node distance=3cm,align=center] (controller) {controller};
    \node [transform_mat, right of=controller, node distance=2.5cm] (m2v) {M2V};
    \node [block, right of=m2v, node distance=3cm] (dm) {DM};
    \node [output, right of=dm, node distance=1.3cm] (output) {};
    %\coordinate  [above of = controller, node distance = 0.8cm]  (A);
    \node [above of = controller, node distance = 0.8 cm] (A)  {RTC};
    \node[fit= (s2m)(m2v) (controller), dashed,draw,inner sep=0.45cm] (Box){};
    % We draw an edge between the controller and dm block to 
    % calculate the coordinate u. We need it to place the measurement block. 
    \draw [->] (controller) -- node[name=u] {$\vec{u}$} (m2v);
    \draw [->] (m2v) -- node[name=v,pos=0.3] {$\vec{v}$} (dm);
    
    \coordinate [below of=u,node distance=1.5cm] (tmp);
    \coordinate [above of=wfs] (tmp2);
    % Once the nodes are placed, connecting them is easy. 
    \draw [draw,->] (input) -- node {$\vec{\phi}$} (sum);
    \draw [->] (sum) -- node [name=e] {$\vec{e}$} (wfs);
    \draw [->] (wfs) --  (sum_noise);
    \draw [->] (sum_noise) -- node[name=s,pos=0.6] {$\vec{s}$} (s2m);
    \draw [->] (s2m) -- node[name=m] {$\vec{m}$} (controller) ;
    \draw [-] (dm) -- node [name=y] {} node [below right] {} (output);
    \draw [->] (output) |- (tmp) -| node[pos=0.99] {$-$} 
        node [near end] {$\vec{\phi}^{corr}$} (sum);
\end{tikzpicture}
}

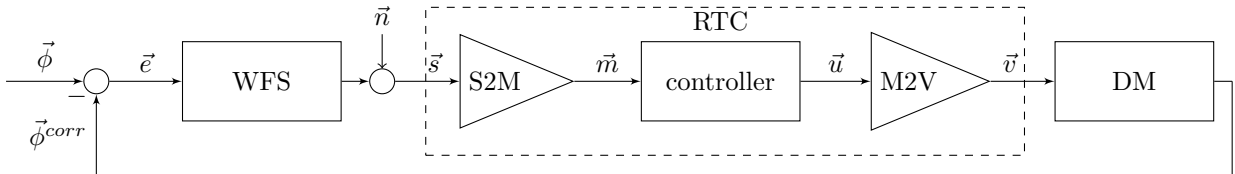
\captionof{figure}{Block diagram of an AO system with modal control. With $\vec{\phi}$ the incoming wavefront,  $\vec{e}$ the residual wavefront, $\vec{n}$ the noise, $\vec{s}$ the slopes, $\vec{m}$ the modal measurements, $\vec{u}$ the modal commands, $\vec{v}$ the voltages, S2M and M2V the modal projection matrices.}
\end{center}
Each of these elements are described with their transfer function. The variable $s$ is used to described the continuous time behaviour (Laplace Transform) and $z$ the discrete time behaviour (Z transform). For simplicity and better readability, vector notation is avoided, as all AO variables are multidimensional.

\subsection{Wavefront sensor}
The wavefront sensor temporal behaviour is an average of the residual $e$ during the integration time $T$
\begin{equation}
    WFS(t) =  \frac{1}{T}\int_{t-T}^{t} e(t) \,dt
\end{equation}
In Laplace transform:
\begin{equation}
    WFS(s) =  \frac{1-e^{-Ts}}{Ts}
\end{equation}
The integration time of the WFS usually defines the loop rate, as RTCs nowadays are capable of reaching higher speeds. The integration time is set depending on the photon flux or the detector capability. In the COMPASS simulation framework, the WFS sensor behaves as a simple delay. Therefore,
\begin{equation}
    WFS(z) =  z^{-1}
\end{equation}
\subsection{Real-time computer}
The real-time computer transfer function is separated in two parts, the inherent computing latency and the control law. 
\begin{equation}
    RTC(s,z) =  e^{-\tau s}K(z)
\end{equation}
with $\tau$ the latency time and $K$ the control law. \(\tau<T\) for the bright cases and \(\tau\approx T\) for the faint cases defined by the consortium. COMPASS handles fractional delay for the RTC by interpolating the commands. In the simulation framework, the latency term encompasses the read-out time, computation time, communication time with the deformable mirror, and the rise time of the latter.

\subsection{Deformable mirror}
No actuators dynamics (except the rise time encompassed in the loop latency) is considered in the simulation framework. Therefore,
\begin{equation}
    DM(s) =  1
\end{equation}
However, the stroke limitation for the deformable mirror is considered. 
\subsection{Modal control}
The data-driven controller uses a Karhunen–Loève modal decomposition of the wavefront such that each mode can be controlled independently. This allows to build multiple Single-Input-Single-Output (SISO) controllers (one for each mode) working in parallel instead of a multi-input-multi-output (MIMO) controller. Mathematically, the wavefront error is decomposed as,
\begin{equation}
\phi(r,\theta,t) = \sum_{j=1}^{\infty} m_j(t)M_j(r,\theta,t)
\label{eqn:wavefront_error}
\end{equation}
with $M_j(r,\theta,t)$ the modal basis and $m_i(t)$ its corresponding coefficient at time $t$. Then any incident wavefront is described by a vector $\vec{m}(t)= m_1(t),m_2(t),..,m_\infty(t)$. In practice a subset of $\vec{m}(t)$ is taken to control a finite number of modes. During calibration the projection matrices from slopes to modes S2M and modes to voltages M2V are computed. 

\subsection{Sensitivity function}
The closed loop transfer function between the disturbance input and the residual is called the sensitivity (or rejection) function and is defined as,
\begin{equation}
\begin{aligned}
    \mathcal{S}(s,z) & =  \frac{1}{1+WFS(s)e^{-\tau s}K(z)DM(s)}\\
   & =  \frac{1}{1+G(s)K(z)}\\
\end{aligned}
\end{equation}
with $G(s)$ the plant.
The complementary sensitivity function is defined as,
\begin{equation}
\begin{aligned}
    \mathcal{T}(s,z) & = \frac{G(s)K(z)}{1+G(s)K(z)}\\
\end{aligned}
\end{equation}

\section{Two stages cascaded AO system}
There are several ways of combining two AO stages. For the SAXO+ baseline two options are considered: standalone and standalone+. The two stages work at different wavelengths and different rates. 
\subsection{Standalone}
The standalone scheme is the simplest form of combining the two stages. The second stage input is the first stage wavefront residual, and no information is shared between the two stages. Tuning the controllers is not trivial in this configuration, as the optimal controller for each of the stages individually may not be the optimal for the whole system. The system is also prone to noise propagation between the two stages. The pros are that the systems benefits from double rejection e.g. if each RTC runs an integrator the whole system acts a an double integrator. 
\begin{center}
\makebox[0pt]{%
\begin{tikzpicture}[auto, node distance=2cm,>=latex']

    % We start by placing the blocks
    \node [input, name=input] {};
    \node [sum, right of=input] (sum) {};
    \node [block, right of=sum,node distance=2cm,align=center] (delay) {$WFS_1$};
    \node [block, right of=delay,node distance=3cm,align=center] (controller) {$RTC_{1}$};
    \node [block, right of=controller, node distance=3cm] (dm) {$DM_{1}$};

    % We draw an edge between the controller and dm block to 
    % calculate the coordinate u. We need it to place the measurement block. 
    \draw [->] (controller) -- node[name=u] {} (dm);
    \node [output, right of=dm] (output) {};
    \coordinate [above of=u,node distance=1.5cm] (tmp);
    \coordinate [right of=dm] (tmp3);
    
    \coordinate [below of=delay, node distance=1.5cm] (tmp4);
    \node [block, right of=tmp4, node distance=0.5cm] (delay2) {$WFS_2$};
    \node [block, right of=delay2,node distance=2.5cm,align=center] (controller2) {$RTC_{2}$};
    \node [block, right of=controller2, node distance=3cm](dm2) {$DM_{2}$};
     
    % Once the nodes are placed, connecting them is easy. 
    \draw [draw,->] (input) -- node {$\phi$} (sum);
    \draw [->] (sum) -- node [name=e] {$e_1$} (delay);
    \node [sum, left of =tmp4, node distance=1.45cm] (sum2) {};
    \draw [->] (e) -- (sum2);
    \draw [->] (sum2) -- node [name=e2] {$e_2$} (delay2);
    \draw [->] (delay) -- (controller);
    %\draw [->] (dm) -- node [name=y] {} node [below right] {$-$} (disturbance);
    \draw [->] (dm) -- (tmp3) |- (tmp) -| node[pos=0.99] {} node [pos=0.95] {$-$} (sum);

    \draw [->] (controller2) -- node[name=u2] {} (dm2);
    \coordinate [below of=u2,node distance=1.5cm] (tmp5);
    \coordinate [right of=dm2] (tmp6);
    \draw [->] (dm2) -- (tmp6) |- (tmp5) -| node[pos=0.99] {} node [pos=0.95] {$-$} (sum2);
    \draw [->] (delay2) -- (controller2);
\end{tikzpicture} 
}

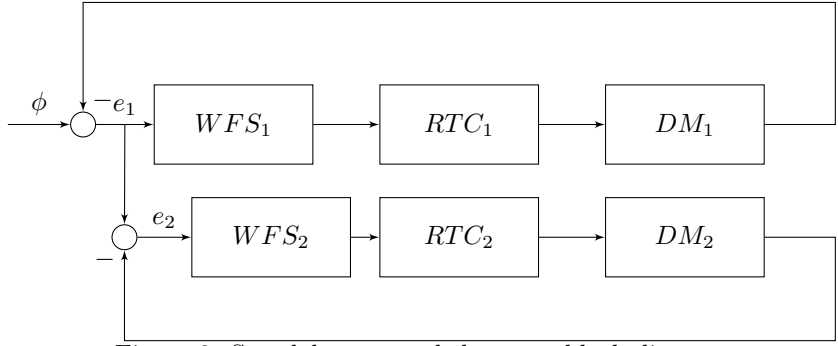
\captionof{figure}{Standalone cascaded system block diagram.}
\end{center}
\subsection{Standalone+}
In the standalone+ configuration the second stage retrieves the first stage commands. This allows in particular to do disentangled Cascaded Adaptive Optics (dCAO) \cite{dcao}. In dCAO mode, the first-stage commands are used to mirror the first-stage DM, ensuring that the correction cancels out and the second stage perceives the full, unaffected disturbance. Compared to the standalone configuration, the system is no longer susceptible to noise propagation, and optimizing each controller individually theoretically maximizes the entire system's performance. However, this approach does not benefit from the double rejection effect.

\begin{center}
\makebox[0pt]{%
\begin{tikzpicture}[auto, node distance=2cm,>=latex']

    % We start by placing the blocks
    \node [input, name=input] {};
    \node [sum, right of=input] (sum) {};
    \node [block, right of=sum,node distance=2cm,align=center] (delay) {$WFS_1$};
    \node [block, right of=delay,node distance=3cm,align=center] (controller) {$RTC_1$};
    \node [block, right of=controller, node distance=3cm] (dm) {$DM_1$};

    % We draw an edge between the controller and dm block to 
    % calculate the coordinate u. We need it to place the measurement block. 
    \draw [->] (controller) -- node[name=u] {} (dm);
    \node  [block, below of=u, node distance=1.7cm](P) {$\mathbf{P}$};
    \node [output, right of=dm] (output) {};
    \coordinate [above of=u,node distance=1.5cm] (tmp);
    \coordinate [right of=dm] (tmp3);
    
    \coordinate [below of=delay, node distance=3cm] (tmp4);
    \node [block, right of=tmp4, node distance=0.5cm] (delay2) {$WFS_2$};
    \node [block, right of=delay2,node distance=2.5cm,align=center] (controller2) {$RTC_2$};
    \node [sum, right of =controller2, node distance=1.5cm] (sum3) {};
    \node [block, right of=sum3, node distance=2cm](dm2) {$DM_2$};
     
    % Once the nodes are placed, connecting them is easy. 
    \draw [draw,->] (input) -- node {$\phi$} (sum);
    \draw [->] (sum) -- node [name=e] {$e_1$} (delay);
    \node [sum, left of =tmp4, node distance=1.45cm] (sum2) {};
    \draw [->] (e) -- (sum2);
    \draw [->] (sum2) -- node [name=e2] {$e_2$} (delay2);
    \draw [->] (delay) -- (controller);
    %\draw [->] (dm) -- node [name=y] {} node [below right] {$-$} (disturbance);
    \draw [->] (dm) -- (tmp3) |- (tmp) -| node[pos=0.99] {} node [pos=0.95] {$-$} (sum);

    \draw [->] (sum3) -- node[name=u2] {} (dm2);
    \draw [->] (controller2) -- (sum3);
    \coordinate [below of=u2,node distance=1.5cm] (tmp5);
    \coordinate [right of=dm2] (tmp6);
    \draw [->] (dm2) -- (tmp6) |- (tmp5) -| node[pos=0.99] {} node [pos=0.95] {$-$} (sum2);
    \draw [->] (delay2) -- (controller2);
    \draw [->] (u) --  (P);
    \draw [->] (P) --  node [pos=0.95]{$-$}(sum3);

\end{tikzpicture} 
}

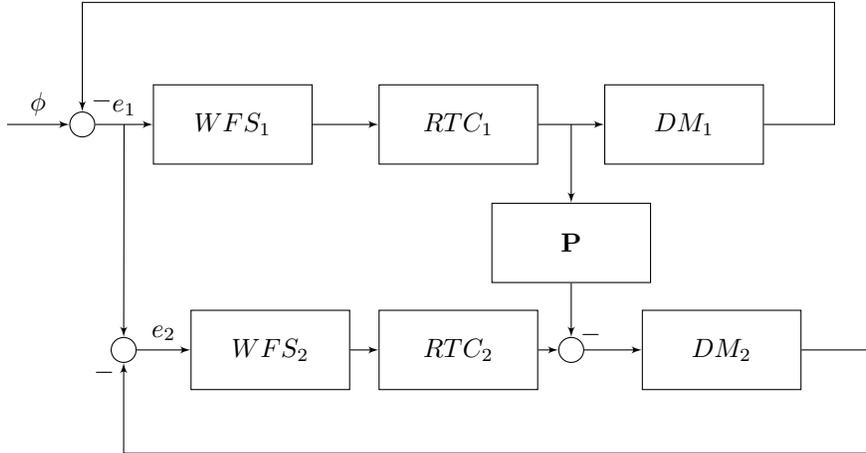
\captionof{figure}{dCAO system block diagram, the matrix $\mathbf{P}$ denotes the projection matrix used to project the first stage commands to the second stage command space.}

\end{center}

\section{Data-driven controller}\label{Data-driven controller}
The data-driven controller is a frequency-based data-driven controller using power spectral
density (PSD) for optimization while ensuring stability criteria. It addresses disturbance rejection, command amplitude constraints and system transfer functions through convex optimization to obtain an optimal control law in an Infinite Input Response (IIR) filter form. Non-parametric identification is used by the optimizer as the optimization is performed on a set of frequency points. The optimization of the data-driven controller and the command computation is agnostic to the cascaded scheme used (i.e. standalone or dCAO).     
In the following subsections the discretized transfer function (z transform) is used to describe each of the transfer functions.
\subsection{Infinite Impulse Response filter}
The data-driven controller is designed as an Infinite Impulse Response (IIR) filter. The equation in the $z$ domain is,
\begin{equation}
\begin{aligned}
    K(z) & = \frac{b_0+b_1z_{-1}+...+b_nz_{-n}}{1+a_1z_{-1}+...+a_nz_{-n}}\\
 K(z) & =  \frac{\sum_{i=0}^n b_iz_{-i}}{1+\sum_{i=1}^n a_iz_{-i}}
\end{aligned}
\end{equation}
where $n$ is the order of the controller.
The IIR filter can be implemented as a digital filter in Direct-Form I. The command at time-step $k$ is computed the following way,
\begin{equation}
\label{eqn:dd_cmd}
    u[k] = \sum_{i=0}^n b_i m[k-i]-\sum_{i=1}^n a_i u[k-i]
\end{equation}
$2n+1$ filter coefficients are to be computed by the optimizer.

\subsection{Optimization objective}
This section describes the objectives and constraints of the optimization problem. The order of the controller is chosen depending on the disturbance spectrum. As the atmosphere has a smooth spectrum, the order can be small. It is eventually increased in presence of vibrations. Keeping the order as low as possible reduces computation time and overfitting.   
\subsubsection{Objective for performance}
The optimizer objective is to minimize the 2-norm of the residual due to the disturbance. The transfer function between the disturbance and the residual is,
\begin{equation}
\begin{aligned}
    e_\phi(z) & = \frac{1}{1+G(z)K(z)}\phi(z)\\
   & =  \mathcal{S}(z)\phi(z)\\
   & = \mathcal{S}(z)\Phi(z)w(z)
\end{aligned}
\end{equation}
where $\Phi(z)$ is the frequency response of the disturbance and $w(z)$ is a white noise signal of magnitude 1.\\
The RMS of the residual induced by the disturbance is,
\begin{equation}
\begin{aligned}
    \norm{e_\phi(z)}_2 & = \norm{\mathcal{S}(z)\Phi(z)w(z)}_2\\
   & =  \norm{\mathcal{S}(z)\Phi(z)}_2\\
\end{aligned}
\end{equation}
\subsubsection{Objective for stroke limitation}
As the actuators have limited stroke, it is essential to minimize the overshoot. The relation between the command induced by the residual and the residual itself is,
\begin{equation}
\begin{aligned}
    u_e(z) & = \frac{WFS(z)RTC(z)}{1+RTC(z)WFS(z)DM(z)}e(z)\\
     & = \frac{WFS(z)RTC(z)DM(z)}{1+RTC(z)WFS(z)DM(z)}e(z)\\
    & =  \mathcal{T}(z)e(z)\\
  \frac{u_e(z)}{e(z)} & = \mathcal{T}(z)
\end{aligned}
\end{equation}
The maximum gain across the frequency domain is,
\begin{equation}
    \norm{\frac{u_e(z)}{e(z)}}_\infty = \norm{\mathcal{T}(z)}_\infty
\end{equation}
\subsubsection{Overall objective}
The overall optimizer objective is,
\begin{equation}
\begin{aligned}
    &\min_{K(z)}\left( \norm{e_\phi(z)}_2^2+ \alpha\norm{\frac{u_e(z)}{\phi(z)}}_\infty^2\right) \\
    &\min_{K(z)}\Big( \norm{\Phi(z)\mathcal{S}(z)}_2^2+ \alpha\norm{\mathcal{T}(z)}_\infty^2\Big) \\
\end{aligned}
\end{equation}
where $\alpha$ is a scalar weight on the command sensitivity objective. The disturbance PSD is normalised with respect to the closed loop bandwidth $\omega_{bw}$.\\
Then the overall objective becomes,
\begin{equation}
\label{eqn:objective}
     \min_{K(z)}\Big( \norm{W_1(z)\mathcal{S}(z)}_2^2+ \norm{W_2(z)\mathcal{T}(z)}_\infty^2\Big) \\
\end{equation}
with,
\begin{equation}
\begin{aligned}
   &W_1(z) = \frac{\Phi(z)}{\Phi(j\omega_{bw})}\\
   &W_2(z) = \alpha
\end{aligned}
\end{equation}

\subsection{$\mathcal{H}_2$ and $\mathcal{H}_\infty$ mixed sensitivity control design}
The mixed sensitivity objective is solved in terms of both $\mathcal{H}_2$ and $\mathcal{H}_\infty$ performance. In a data-driven approach the problem is solved for all frequency points. Note that in the following, the argument $z$ is omitted to simplify the notation. Equation \ref{eqn:objective} is rewritten as, 
\begin{equation}
    \min_{K}\left(\gamma+\int_{-\frac{\pi}{T_s}}^{\frac{\pi}{T_s}}\textrm{trace}[\Gamma(\omega)]d\omega\right)
\end{equation}
subject to,
\begin{equation}
\label{eqn:cons1}
 (W_1\mathcal{S})(W_1\mathcal{S})^*\prec\Gamma(\omega) \hspace{1cm} \forall\omega \in \Omega \\
\end{equation}
and,
\begin{equation}
\label{eqn:cons2}
(W_2\mathcal{T})(W_2\mathcal{T})^*\prec\gamma I \hspace{1.3cm} \forall\omega \in \Omega \\
\end{equation}
where $\Omega$ is a set of frequencies between $0$ and the Nyquist frequency, $f_s\pi$. $\Gamma(\omega)$ and $\gamma$ are optimization variables, with $\Gamma(\omega)$ changing over the frequencies. \\
The constraints are transformed into a set of LMIs in order to be solved with a convex optimization solver.\\
For constraint \ref{eqn:cons1}, by replacing $K = XY^{-1}$ and $\mathcal{S}=(I+GK)^{-1}$,
\begin{equation}
W_1Y(Y+GX)^{-1}[(Y+GX)^{-1}]^*(W_1Y)^*\prec\Gamma(\omega)\\
\end{equation}
Then taking $P = Y+GX$,
\begin{equation}
\Gamma(\omega)-W_1Y(P^{*}P)^{-1}(W_1Y)^*\succ 0\\
\end{equation}
Which is a convex-concave problem and can be convexified using Quadratic Matrix Inequality (QMI) Convexification Lemma, and choosing $P_c = Y_c + GX_c$, where $K_c = X_cY^{-1}_c$ is the initial stabilizing controller,
\begin{equation}
\begin{bmatrix}
\Gamma(\omega) & W_1Y\\
(W_1Y)^* & P^*P_c+P^*_cP-P^*_cP_c
\end{bmatrix}
\succ 0
\end{equation}
Analogously constraint \ref{eqn:cons2} becomes through convexification,
\begin{equation}
\begin{bmatrix}
\gamma I & W_2GX\\
(W_2GX)^* & P^*P_c+P^*_cP-P^*_cP_c
\end{bmatrix}
\succ 0
\end{equation}
To sum up the convex problem is,
\begin{equation}
    \min_{X,Y}\left(\gamma+\int_{-\frac{\pi}{T_s}}^{\frac{\pi}{T_s}}\textrm{trace}[\Gamma(\omega)]d\omega\right)
\end{equation}
subject to,
\begin{equation}
\begin{bmatrix}
\Gamma(\omega) & W_1Y\\
(W_1Y)^* & P^*P_c+P^*_cP-P^*_cP_c
\end{bmatrix}
\succ 0 \hspace{0.25cm}\textrm{and}\hspace{0.25cm}
\begin{bmatrix}
\gamma I & W_2GX\\
(W_2GX)^* & P^*P_c+P^*_cP-P^*_cP_c
\end{bmatrix}
\succ 0 \hspace{0.5cm} \forall\omega \in \Omega
\end{equation}
\subsection{Robust stability}
For robust stability a modulus margin $\mu$ of the closed system is set, which translates as the following constraint for the optimizer,
\begin{equation}
\norm{\mathcal{S}}_\infty <\mu^{-1} \Rightarrow (\mu\mathcal{S})(\mu\mathcal{S})^*\prec I \hspace{1.3cm} \forall\omega \in \Omega \\
\end{equation}
in the linear matrix linearity form,
\begin{equation}
\begin{bmatrix}
I & \mu Y\\
(\mu Y)^* & P^*P_c+P^*_cP-P^*_cP_c
\end{bmatrix}
\succ 0
\end{equation}
\section{Noise and disturbances}
This sections describes noise and disturbances sources considered in the simulations. 
\subsection{Noise}

Two noise sources are considered in the simulation framework: the flux noise and the readout noise.
The signal-to-noise ratio (SNR) is defined as the ratio of mean to standard deviation of the signal,
\begin{equation}
    \textrm{SNR} = \frac{N_q}{\sqrt{\sigma_\phi^2+\sigma_{N_r}^2}}
\end{equation} 
where, $N_q$ it the number of incident photons per timestep, $\sigma_\phi^2$ is the variance due to flux noise, and $\sigma_{N_r}^2$is the variance due to sensor read-out noise.\\
\subsubsection{Flux noise}
The flux noise can be understood as the number of photons reaching the detecting cell. If we view a photon as a discrete particle, the arrival of photons on the detector surface can be modeled as a Poisson process, as photon arrivals are independent from each other. Consequently, there exists a variation in intensity between two frames. As per the Poisson distribution, the variance equals the expected value of incident photons,
\begin{equation}
    \sigma_\phi^2 = N_q
\end{equation}  
\subsubsection{Readout noise}
The read-out noise arises from electrons that are not generated by an incident photon. This phenomenon is intrinsic to the physical processes within a CCD, and the number of extra electrons depends on the CCD's operating temperature.  The variance due to the readout noise is given by,
\begin{equation}
    \sigma_{N_r}^2 = 4N_r^2
\end{equation}

\subsection{Disturbance}
The main disturbance source is the atmospheric turbulence. Changes in temperature and wind speed between
different atmospheric layers cause the light to diffract and distorts wavefront. Other sources are the telescope wind shake, instrument vibrations, low-wind effect, etc. 
\subsubsection{Atmospheric turbulence}
For the consolidation review, only the atmospheric turbulence is considered as disturbance source. The atmospheric turbulence is simulated in COMPASS as described in F. Assémat and al. 2006\cite{atm}. A 35 layers wind driven halo profile provided by the ESO \cite{turb} is used. The science cases specify the Fried parameter $r_0$, the coherence time $\tau_o$ and the photon flux. 
\subsubsection{Vibration}\label{Vibration}
For an additional scenario, vibrations are also taken into account. Integrator controllers typically do not handle vibrations well, but advanced controllers do. A VLT vibration model is constructed using pseudo open-loop data obtained with the SPHERE instrument. Initially, the atmospheric effects are filtered out, and then the vibration is modeled by fitting a finite impulse response (FIR) filter. This filter enables the generation of vibrations on the fly by convolving with white noise.
\begin{figure}[H] 
\begin{minipage}[b]{0.5\linewidth}
\centering
\includegraphics[width=8cm]{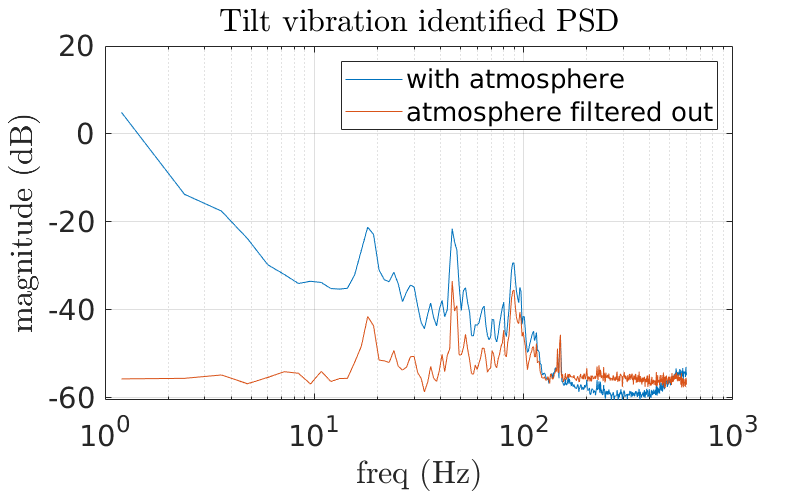}
\captionsetup{width=8cm}
\caption{Tilt vibration power spectral density after filtering out the atmospheric disturbance.}
\label{fig:rms_r0}
% \vspace{1ex}
\end{minipage}%%
\begin{minipage}[b]{0.5\linewidth}
\centering

\includegraphics[width=8cm]{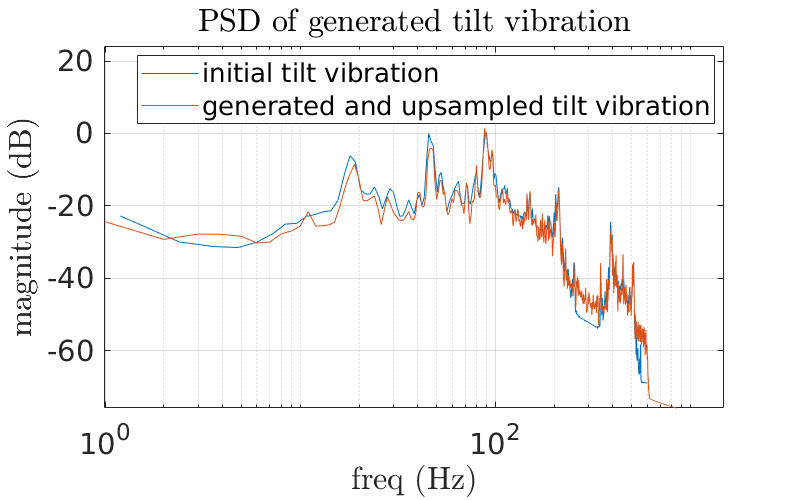}
\captionsetup{width=8cm}
\caption{Tilt vibration power spectral density obtained by filtering the model with white noise.}
\label{fig:psd_r0}
\end{minipage} 
\end{figure}

\section{Science cases results}
Five science cases are assessed for the consolidation review. For each stage, the frequency is optimized using the analytical tool PAOLA \cite{PAOLA}. The number of controlled modes remains consistent across different control strategies, determined empirically using a standard integrator. The reference controller is an integrator with a single modal gain, hand-tuned to achieve the best contrast between 4 and 6 $\lambda/D$. This control remains unchanged for the first stage. For the second stage, it is compared against a data-driven controller with an integrator structure, where the gain of the integrator is optimized for each mode, as described in section \ref{Data-driven controller}, referred as optimized modal gain integrator in the following. This setup allows for cross-checking with the hand-tuned integrator. The third control strategy employs a fifth-order data-driven controller, simply referred as data-driven in the following. Each science case is evaluated in standalone and dCAO mode. The performance metrics used include the average contrast after a perfect coronagraph between 4 and 6 $\lambda/D$, the Strehl ratio evaluated at 1.65 \si{\micro\meter} and the radial average contrast curve. The simulations exposure time is set to 5 seconds of stationary atmospheric condition.

\subsection{Bright 1 best}\label{bright1best}
This first case represents the most optimistic observing conditions: high photon flux in both G and J band, low seeing and low wind speed. 
\begin{table}[H]
\caption{Bright 1 best, case parameters.}
\centering
\begin{tabular}{||c | c | c | c | c | c | c | c | c | c||} 
 \hline
 seeing & $t_0$ & G mag & J mag & \makecell{\nth{1} st. \\ frequency} & \makecell{\nth{2} st. \\ frequency} & \makecell{\nth{1} st. \\ \texttt{\#} of modes} & \makecell{\nth{2} st. \\ \texttt{\#} of modes} & \makecell{\nth{1} st. \\ $\lambda$} & \makecell{\nth{2} st. \\ $\lambda$}\\ [0.5ex] 
 \hline
 0.4" & 9 ms & 5.5 & 5.2 & 1000 Hz & 3000 Hz & 1200 & 400 & 0.69 \si{\micro\meter} &  1.04 \si{\micro\meter}\\ 
 \hline
\end{tabular}
\end{table}

\begin{figure} [H]
\begin{center}
\begin{tabular}{c} %% tabular useful for creating an array of images 
\includegraphics[height=7cm]{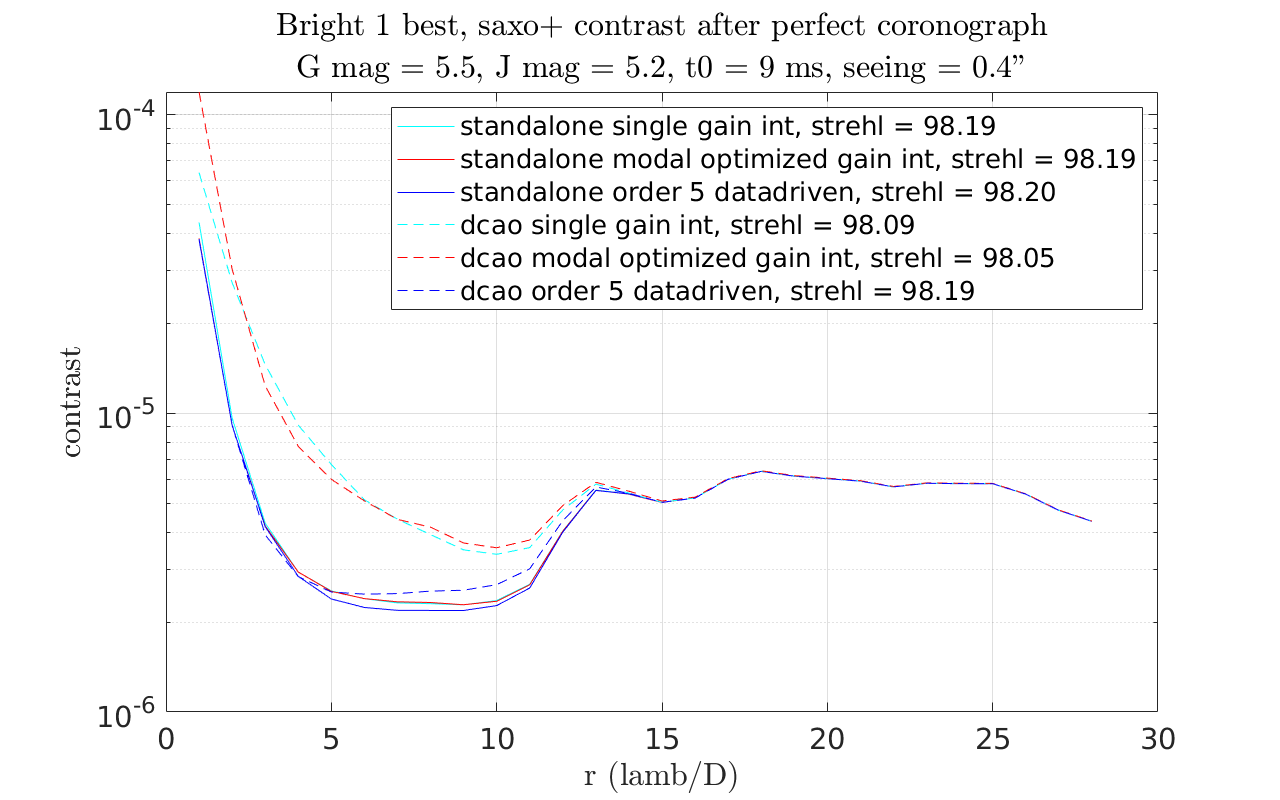}
\end{tabular}
\end{center}
\caption[example] 
%>>>> use \label inside caption to get Fig. number with \ref{}
{ \label{fig:example} 
Radial average contrast comparison between the control strategies for the "bright 1 best" case.}
\end{figure} 

\begin{table}[H]
\caption{Bright 1 best, performance comparison.}
\centering
\begin{tabular}{||c | c | c | c | c | c | c||} 
 \hline
  & \makecell{standalone \\ reference} & \makecell{standalone \\ opt. modal gain} & \makecell{standalone \\ data-driven} & \makecell{dCAO \\ reference} & \makecell{dCAO \\ opt. modal gain} & \makecell{dCAO \\ data-driven} \\ [0.5ex] 
 \hline
 \makecell{Avg contrast \\ 4-6 $\lambda/D$} & 2.62e-6 & 2.62e-6 & \textcolor{red}{2.49e-6} & 6.99e-6 & 6.28e-6 & 2.61e-6 \\ 
  \hline
 strehl [\verb|%|] & 98.19 & 98.19 & \textcolor{red}{98.20} & 98.09 & 98.05 & 98.19 \\ 
 \hline
\end{tabular}
\end{table}

\subsection{Bright 1 worst}
This case keeps the same photon flux as previous one, however, with worse seeing and higher wind speed. 
\begin{table}[H]
\caption{Bright 1 worst, case parameters.}
\centering
\begin{tabular}{||c | c | c | c | c | c | c | c | c | c||} 
 \hline
 seeing & $t_0$ & G mag & J mag & \makecell{\nth{1} st. \\ frequency} & \makecell{\nth{2} st. \\ frequency} & \makecell{\nth{1} st. \\ \texttt{\#} of modes} & \makecell{\nth{2} st. \\ \texttt{\#} of modes} & \makecell{\nth{1} st. \\ $\lambda$} & \makecell{\nth{2} st. \\ $\lambda$}\\ [0.5ex] 
 \hline
 1.0" & 2 ms & 5.5 & 5.2 & 1000 Hz & 3000 Hz & 1200 & 540 & 0.69 \si{\micro\meter} &  1.04 \si{\micro\meter}\\ 
 \hline
\end{tabular}
\end{table}

\begin{figure} [H]
\begin{center}
\begin{tabular}{c} %% tabular useful for creating an array of images 
\includegraphics[height=7cm]{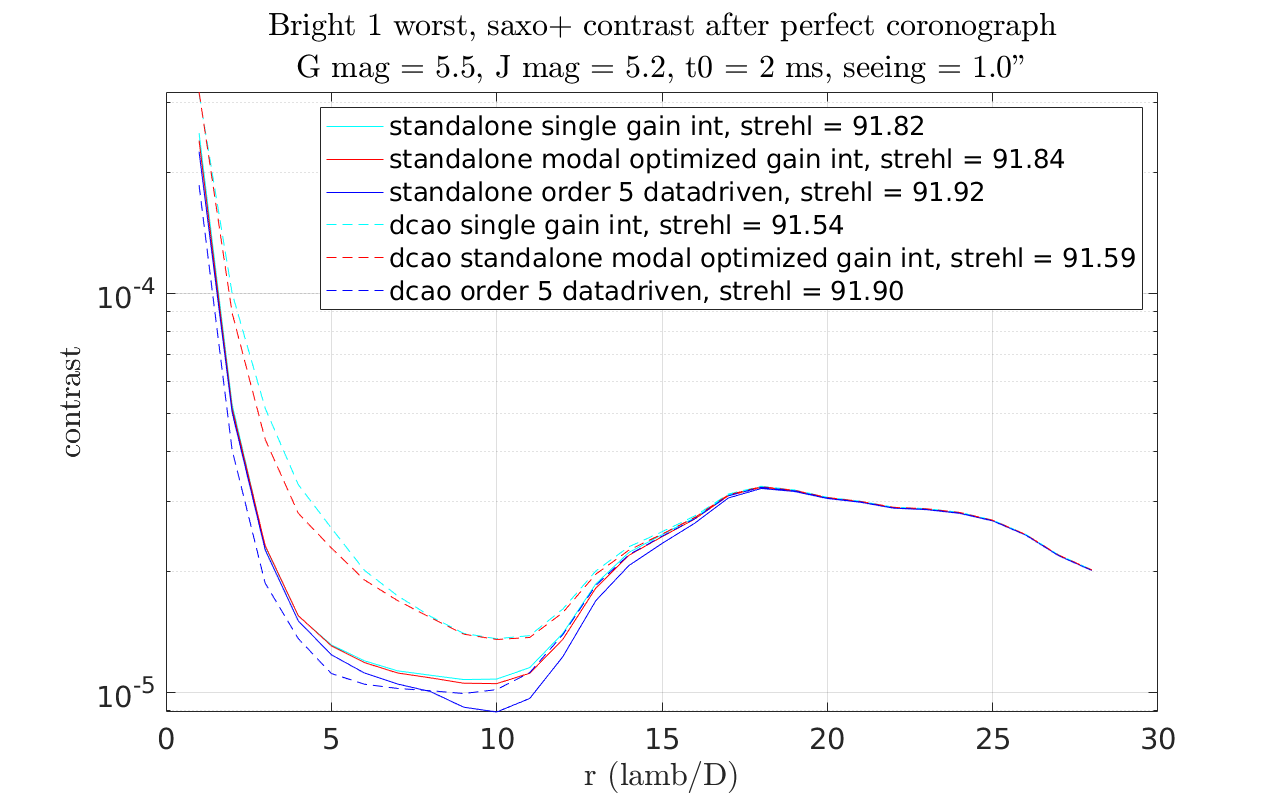}
\end{tabular}
\end{center}
\caption[example] 
%>>>> use \label inside caption to get Fig. number with \ref{}
{ \label{fig:example} 
Radial average contrast comparison between the control strategies for the "bright 1 worst" case.}
\end{figure} 

\begin{table}[H]
\caption{Bright 1 worst, performance comparison.}
\centering
\begin{tabular}{||c | c | c | c | c | c | c||} 
 \hline
  & \makecell{standalone \\ reference} & \makecell{standalone \\ opt. modal gain} & \makecell{standalone \\ data-driven} & \makecell{dCAO \\ reference} & \makecell{dCAO \\ opt. modal gain} & \makecell{dCAO \\ data-driven} \\ [0.5ex] 
 \hline
 \makecell{Avg contrast \\ 4-6 $\lambda/D$} & 1.35e-5 & 1.35e-5 & 1.29e-5 & 2.64e-5 & 2.39e-5 & \textcolor{red}{1.17e-5} \\ 
  \hline
 strehl [\verb|%|] & 91.82 & 91.84 & \textcolor{red}{91.92} & 91.54 & 91.59 & 91.90 \\ 
 \hline
\end{tabular}
\end{table}

\subsection{Red 1 fast}
The "red" cases are fainter and have a higher flux in the J band than in the G one. Therefore, the first stage (SPHERE) receives a very limited number of photons. "Red 1" is the brightest of the "red" cases. It has a median seeing and fast wind speed.   
\begin{table}[H]
\caption{Red 1 fast, case parameters.}
\centering
\begin{tabular}{||c | c | c | c | c | c | c | c | c | c||} 
 \hline
 seeing & $t_0$ & G mag & J mag & \makecell{\nth{1} st. \\ frequency} & \makecell{\nth{2} st. \\ frequency} & \makecell{\nth{1} st. \\ \texttt{\#} of modes} & \makecell{\nth{2} st. \\ \texttt{\#} of modes} & \makecell{\nth{1} st. \\ $\lambda$} & \makecell{\nth{2} st. \\ $\lambda$}\\ [0.5ex] 
 \hline
 0.7" & 2 ms & 11.9 & 8.5 & 300 Hz & 3000 Hz & 400 & 540 & 0.69 \si{\micro\meter} &  1.04 \si{\micro\meter} \\ 
\hline
\end{tabular}
\end{table}

\begin{figure} [H]
\begin{center}
\begin{tabular}{c} %% tabular useful for creating an array of images 
\includegraphics[height=7cm]{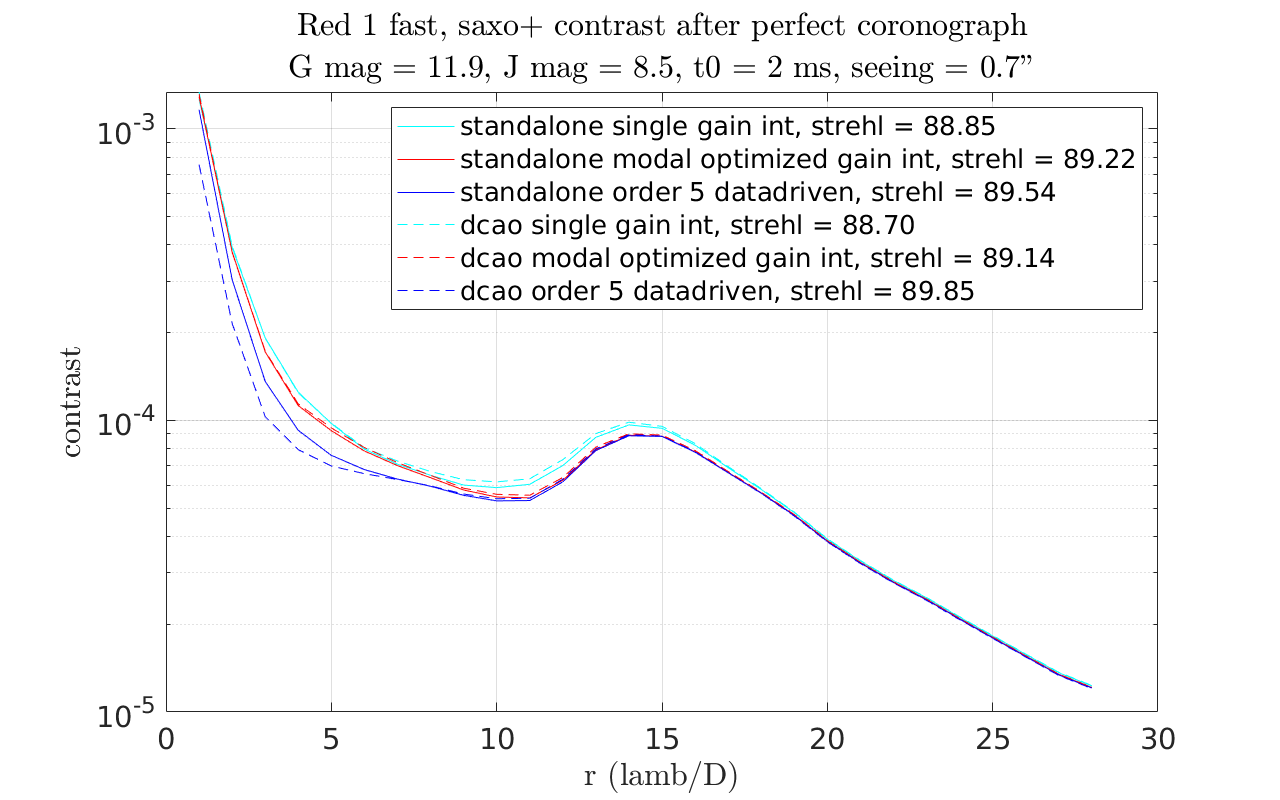}
\end{tabular}
\end{center}
\caption[example] 
%>>>> use \label inside caption to get Fig. number with \ref{}
{ \label{fig:example} 
Radial average contrast comparison between the control strategies for the "red 1 fast" case.}
\end{figure} 

\begin{table}[H]
\caption{Red 1 fast, performance comparison.}
\centering
\begin{tabular}{||c | c | c | c | c | c | c||} 
 \hline
  & \makecell{standalone \\ reference} & \makecell{standalone \\ opt. modal gain} & \makecell{standalone \\ data-driven} & \makecell{dCAO \\ reference} & \makecell{dCAO \\ opt. modal gain} & \makecell{dCAO \\ data-driven} \\ [0.5ex] 
 \hline
 \makecell{Avg contrast \\ 4-6 $\lambda/D$} & 1.00e-4 & 9.44e-5 & 7.87e-5 & 1.00e-4 & 9.63e-5 & \textcolor{red}{7.16e-5} \\ 
  \hline
 strehl [\verb|%|] & 88.85 & 89.22 & 89.54 & 88.70 & 89.14 & \textcolor{red}{89.85} \\ 
 \hline
\end{tabular}
\end{table}

\subsection{Red 3 medium}
"Red 3" is a fainter "red" case. A median seeing and average wind speed is chosen. 
\begin{table}[H]
\caption{Red 3 medium, case parameters.}
\centering
\begin{tabular}{||c | c | c | c | c | c | c | c | c | c||} 
 \hline
 seeing & $t_0$ & G mag & J mag & \makecell{\nth{1} st. \\ frequency} & \makecell{\nth{2} st. \\ frequency} & \makecell{\nth{1} st. \\ \texttt{\#} of modes} & \makecell{\nth{2} st. \\ \texttt{\#} of modes} & \makecell{\nth{1} st. \\ $\lambda$} & \makecell{\nth{2} st. \\ $\lambda$}\\ [0.5ex] 
 \hline
 0.7" & 5.5 ms & 14.5 & 10.1 & 50 Hz & 1250 Hz & 400 & 540 & 0.69 \si{\micro\meter} &  1.14 \si{\micro\meter} \\ 
\hline
\end{tabular}
\end{table}

\begin{figure} [H]
\begin{center}
\begin{tabular}{c} %% tabular useful for creating an array of images 
\includegraphics[height=7cm]{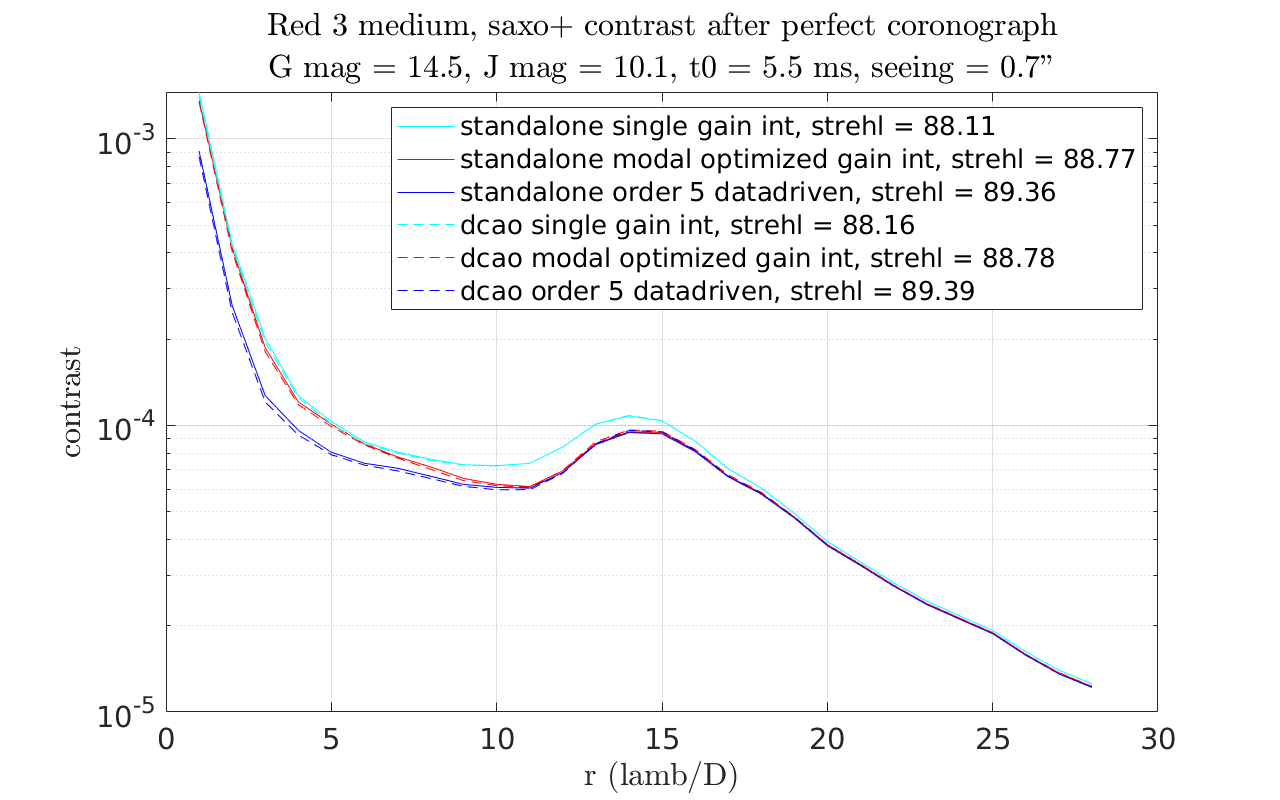}
\end{tabular}
\end{center}
\caption[example] 
%>>>> use \label inside caption to get Fig. number with \ref{}
{ \label{fig:example} 
Radial average contrast comparison between the control strategies for the "red 3 medium" case.}
\end{figure} 

\begin{table}[H]
\caption{Red 3 medium, performance comparison.}
\centering
\begin{tabular}{||c | c | c | c | c | c | c||} 
 \hline
  & \makecell{standalone \\ reference} & \makecell{standalone \\ opt. modal gain} & \makecell{standalone \\ data-driven} & \makecell{dCAO \\ reference} & \makecell{dCAO \\ opt. modal gain} & \makecell{dCAO \\ data-driven} \\ [0.5ex] 
 \hline
 \makecell{Avg contrast \\ 4-6 $\lambda/D$} & 1.06e-4 & 1.03e-4 & 8.35e-5 & 1.04e-4 & 1.01e-4 & \textcolor{red}{8.15e-5} \\ 
  \hline
 strehl [\verb|%|] & 88.11 & 88.77 & 89.36 & 88.16 & 88.78 & \textcolor{red}{89.39} \\ 
 \hline
\end{tabular}
\end{table}

\subsection{Red 5 best}
"Red 5" is the faintest case considered. Best atmospheric conditions are set with a low seeing and low wind speed. 
\begin{table}[H]
\caption{Red 5 best, case parameters.}
\centering
\begin{tabular}{||c | c | c | c | c | c | c | c | c | c||} 
 \hline
 seeing & $t_0$ & G mag & J mag & \makecell{\nth{1} st. \\ frequency} & \makecell{\nth{2} st. \\ frequency} & \makecell{\nth{1} st. \\ \texttt{\#} of modes} & \makecell{\nth{2} st. \\ \texttt{\#} of modes} & \makecell{\nth{1} st. \\ $\lambda$} & \makecell{\nth{2} st. \\ $\lambda$}\\ [0.5ex] 
 \hline
 0.4" & 9 ms & 16.8 & 12.5 & 10 Hz & 300 Hz & 400 & 400 & 0.69 \si{\micro\meter} &  1.14 \si{\micro\meter} \\ 
\hline
\end{tabular}
\end{table}

\begin{figure} [H]
\begin{center}
\begin{tabular}{c} %% tabular useful for creating an array of images 
\includegraphics[height=7cm]{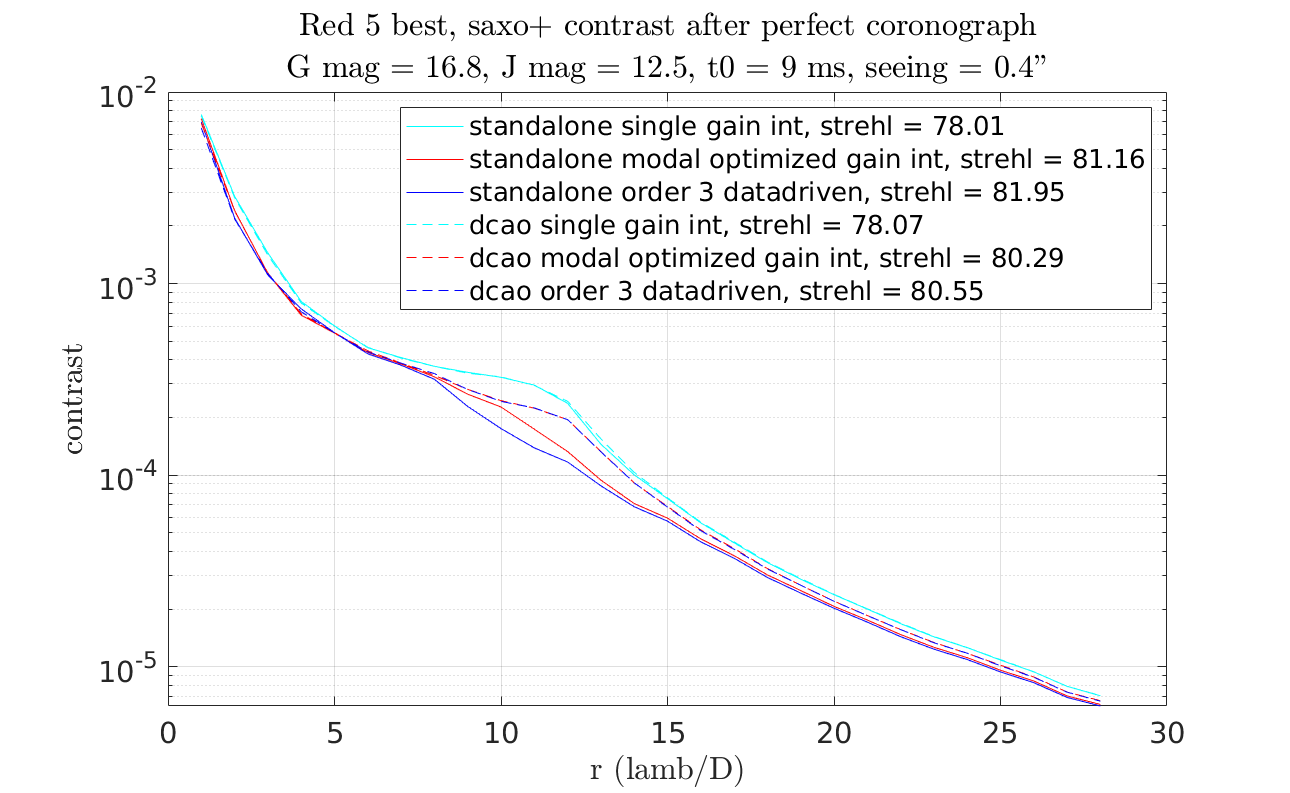}
\end{tabular}
\end{center}
\caption[example] 
%>>>> use \label inside caption to get Fig. number with \ref{}
{ \label{fig:example} 
Radial average contrast comparison between the control strategies for the "red 5 best" case.}
\end{figure} 

\begin{table}[H]
\caption{Red 5 best, performance comparison.}
\centering
\begin{tabular}{||c | c | c | c | c | c | c||} 
 \hline
  & \makecell{standalone \\ reference} & \makecell{standalone \\ opt. modal gain} & \makecell{standalone \\ data-driven} & \makecell{dCAO \\ reference} & \makecell{dCAO \\ opt. modal gain} & \makecell{dCAO \\ data-driven} \\ [0.5ex] 
 \hline
 \makecell{Avg contrast \\ 4-6 $\lambda/D$} & 6.23e-4 & \textcolor{red}{5.58e-4}& 5.74e-4 & 6.18e-4 & 5.65e-4 & 5.69e-4 \\ 
  \hline
 strehl [\verb|%|] & 78.01 & 81.16 & \textcolor{red}{81.95} & 78.07 & 80.29 & 80.55 \\ 
 \hline
\end{tabular}
\end{table}

\subsection{Comments on results}
\subsubsection{Crosscheck between single hand-tuned gain and optimized modal gain integrator}
The single hand-tuned gain and optimized modal gain integrator exhibit similar performance, thereby cross-validating both strategies. In the "red" cases, the optimized modal gain integrator demonstrates an enhancement in contrast at higher radius ($>6\lambda/D$). In these fainter cases, the optimized modal gain calculates a lower gain for the high order modes. 

\subsubsection{Data-driven controller}
The data-driven controller exhibits superior performance compared to the integrator, particularly in the presence of fast wind, where the disturbance spectrum shifts towards higher frequencies. Consequently, a first-order low-pass filter like the integrator becomes sub-optimal. However, the improvement may not be significant enough to be noticeable in the presence of other error sources that are not considered, such as non-common path aberrations (NCPAs).

\subsubsection{Standalone vs. dCAO}
The distinction between standalone and dCAO is more pronounced in the "bright" cases. The performance disparity stemming from the double rejection effect is significant, but it is mitigated when employing a higher-order controller like the data-driven one. In such instances, the performance is comparable between the two strategies. The advantage of using dCAO lies not in performance improvement, but rather in the separation of both stages, which simplifies the tuning of controllers. In the fainter "red" cases, the first stage operates slower with a low gain mitigating its impact on performance. Therefore, disentangling both stages does no show great difference. 

\section{Performance under vibrations}
The primary strength of advanced control strategies lies in their ability to adapt and reject disturbance sources such as vibrations. This capability is demonstrated by comparing the sensitivity function of the data-driven controller with that of an integrator in the presence of vibrations. In this experiment, the tilt vibration described in subsection \ref{Vibration}  is added to the atmospheric disturbance of the first science case, "bright 1 best" (\ref{bright1best}). For the first stage, an integrator is used, while for the second stage, the integrator is compared to the data-driven controller. The data-driven optimizer places notches to damp the main vibration peaks, resulting in a flatter residual spectrum and, consequently, a lower residual RMS.
 
\begin{figure}[H] 
\begin{minipage}[b]{0.5\linewidth}
\centering
\includegraphics[width=8cm]{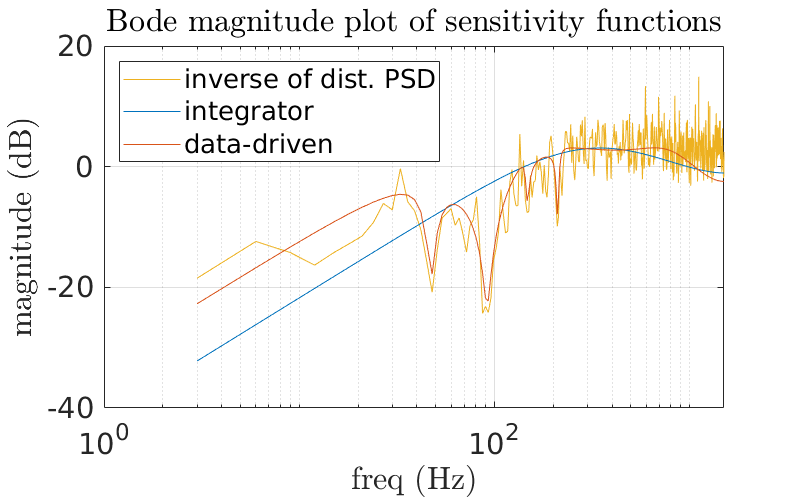}
\captionsetup{width=8cm}
\caption{Sensitivity function comparison between an integrator and a data-driven controller along with the disturbance at the input of the second stage containing vibrations.}
\label{fig:rms_r0}
% \vspace{1ex}
\end{minipage}%%
\begin{minipage}[b]{0.5\linewidth}
\centering

\includegraphics[width=8cm]{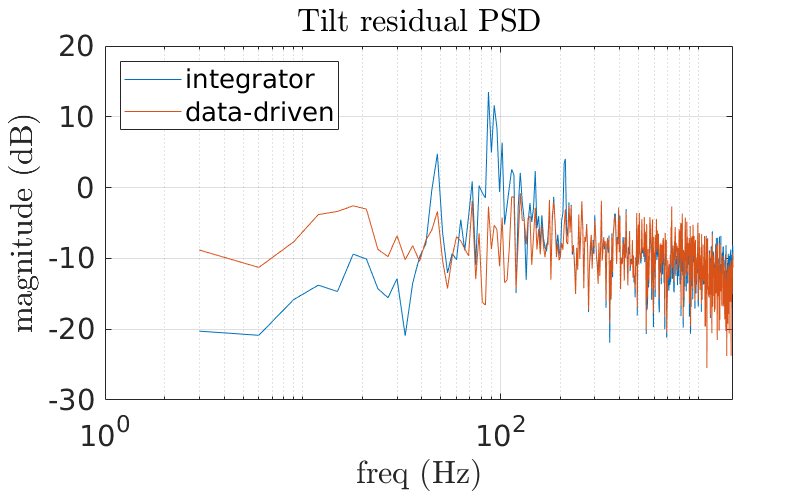}
\captionsetup{width=8cm}
\caption{Tilt residual power spectral density comparison between the integrator and data-driven controller in presence of vibrations. The data-driven shows a flatter spectrum.}
\label{fig:psd_r0}
\end{minipage} 
\end{figure}

\begin{table}[H]
\caption{Tilt residual RMS comparison in presence of vibrations.}
\centering
\begin{tabular}{||c | c | c ||} 
 \hline
integrator & data-driven & relative gain \\ [0.5ex] 
 \hline
 9.09 nm & 6.21 nm  & 31.7 [\verb|%|]\\ 
 
 \hline
\end{tabular}
\end{table}
\section{Conclusion}

%In conclusion, the data-driven controller demonstrated its capability in disturbance rejection in a two-stage AO system. Further work will involve testing it with non-stationary atmospheric conditions to determine the update rate of the filter coefficients. Testing the data-driven controller on a test bed and in the sky is also planned. Regarding the dCAO scheme, while it may offer some advantages, it should not be used a priori with a first-order integrator controller. Further studies will be conducted on this subject as well.
In conclusion, the data-driven controller has showcased its proficiency in disturbance rejection within a two-stage AO system. Further work will involve testing it with non-stationary atmospheric conditions to determine the update rate of the filter coefficients. It has been tested and validated on bench and we look forward to test it on sky within the SAXO+ project. As for the dCAO scheme, while it potentially offers advantages, it should not be automatically used with a first-order integrator controller. Further investigations will be undertaken to delve deeper into this matter.

\acknowledgments % equivalent to \section*{ACKNOWLEDGMENTS}       
 
SAXO+ is an instrument designed and built by a consortium consisting of LESIA, CRAL, INAF, IPAG, Lagrange, LCF, University of Geneva, MPIA, and LAM, in collaboration with ESO. The SAXO+ project received funding from The Fondation Charles Defforey-Institut de France, CNRS/INSU, the Programme et Equipements Prioritaires de Recherche “Origins”, the Piano Nazionale di Ripresa e Resilienza, PlanetS, and MPIA.
\newpage
% References
\bibliography{report} % bibliography data in report.bib
\bibliographystyle{spiebib} % makes bibtex use spiebib.bst

\end{document}